\documentclass[a4paper,11pt]{article}

\usepackage{jinstpub_edit} 
\usepackage{tabularx}
\usepackage{booktabs}
\usepackage{multirow}
\usepackage{array}
\usepackage{gensymb}

\usepackage{caption}
\usepackage{subcaption}
\usepackage[dvipsnames]{xcolor}

\usepackage{lineno}
\title{\boldmath Quantitative Comparison of Planar Coded Aperture Imaging Reconstruction Methods}

\author[a,b,1]{Mei{\ss}ner, T.\note{Corresponding author.}}
\author[c]{Rozhkov, V.}
\author[a, d, e, f]{Hesser, J.}
\author[b]{Nahm, W.}
\author[a]{Loew, N.}

\affiliation[a]{Mannheim Institute for Intelligent Systems in Medicine (MIISM), Medical Faculty Mannheim, Heidelberg University, Mannheim, Germany}
\affiliation[b]{Institute of Biomedical Engineering (IBT), Karlsruhe Institute of Technology (KIT), Karlsruhe, Germany}
\affiliation[c]{International Intergovernmental Organization Joint Institute for Nuclear Research, Dubna, Russia}
\affiliation[d]{Interdisciplinary Center for Scientific Computing (IWR), Heidelberg University, Heidelberg, Germany}
\affiliation[e]{Central Institute for Computer Engineering (ZITI), Heidelberg University, Heidelberg, Germany}
\affiliation[f]{CZS Heidelberg Center for Model-Based AI, Heidelberg University, Mannheim, Germany}

\emailAdd{tobias.meissner@medma.uni-heidelberg.de}

\abstract{
Imaging distributions of radioactive sources plays a substantial role in nuclear medicine as well as in monitoring nuclear waste and its deposit. Coded Aperture Imaging has been proposed as an alternative to parallel or pinhole collimators, but requires image reconstruction as an extra step. Multiple reconstruction methods with varying run time and computational complexity have been proposed. Yet, no quantitative comparison between the different reconstruction methods has been carried out so far. 
This paper focuses on a comparison based on three sets of hot-rod phantom images captured with an experimental $\gamma$-camera consisting of a Tungsten-based MURA mask with a 2\,mm thick 256\,$\times$\,256 pixelated CdTe semiconductor detector coupled to a Timepix\textsuperscript{\textcopyright} readout circuit.
Analytical reconstruction methods, MURA Decoding, Wiener Filter and a convolutional Maximum Likelihood Expectation Maximization (MLEM) algorithm were compared to data-driven Convolutional Encoder-Decoder (CED) approaches. The comparison is based on the contrast-to-noise ratio as it has been previously used to assess reconstruction quality. For the given set-up, MURA Decoding, the most commonly used CAI reconstruction method, provides robust reconstructions despite the assumption of a linear model. For single image reconstruction, however, MLEM performed best among analytical reconstruction methods, but took the longest with an average of 13\,s run time. The fastest reconstruction method is the Wiener Filter with 67\,ms and mediocre quality. 
The CED with a specifically tailored training set was able to succeed the most commonly used MURA decoding on average by a factor between 1.37 and 2.60 and a run time of around 300\,ms.
}


\begin{document}
\maketitle
\flushbottom

\section{Introduction}
\label{sec:intro}
Accurate localization and comprehensible visualization of radioactive source distributions constitutes the essence of nuclear medicine \cite{Tsuchimochi2013, Matthies2013, VanAudenhaege2015, Kennedy2014, Peterson2011, Roth2020, Fujii2012}, nuclear waste monitoring \cite{Gal2006, Gmar2011, vetter2018gamma, Cieslak2016, Sun2020} and x-ray fluorescence microscopy \cite{Belthangady2019, Kulow2020}. When high-energy photons are involved, refractive or reflective camera systems are not realizable due to the negative refraction index or demand expensive manufacturing processes due to the high surface quality for x-ray mirrors. 
Thus, modern $\gamma$-cameras are equipped with pinhole or parallel collimators to acquire the spatial information of incident rays. This kind of collimation represents a trade-off between resolution and photon efficiency: The smaller the openings or the higher the aspect ratio, the sharper the image, but the less photons pass through the collimator \cite{Madsen2007, Tsuchimochi2013, Peterson2011}.
Due to this limitation, Coded Aperture Imaging (CAI) has been discussed, because it persuades a better compromise between resolution and photon efficiency. Proposed independently by Ables \cite{Ables1968} and Dicke \cite{dicke1968scatter} for X- and $\gamma$-ray astronomy (see \cite{Caroli1987, Braga2020} for further information), a mask between object and detector consisting of a radiopaque material with transparent elements in between, encodes direction information of incoming $\gamma$-rays. Since each transparent element of the mask acts as a pinhole and thus generates a projection of the object on the detector, the resulting image consists of a multitude of overlapping images and becomes incomprehensible. Therefore, image reconstruction is required to obtain an interpretable image.\\
CAI has been investigated as alternative collimator in single photon emission computed tomography (SPECT): Experiments for imaging the bio-distribution of radioactively labeled compounds in small animals, have been carried out \cite{Accorsi2001a, Accorsi2007}. \cite{Fujii2012, Russo2020, Kaissas2015} analyzed the use of a coded aperture $\gamma$-camera for the localization of sentinel lymph nodes and \cite{Mu2006, Mu2016} examined the capability of imaging cold-spot lesions in cardiac imaging. 
Other research fields like x-ray fluorescence spectroscopy \cite{Kulow2020} or nuclear decommissioning \cite{Cieslak2016, Gmar2011} analyzed the potential use of CAI as well. \\
When the object’s longitudinal extension is small in comparison to the camera-object distance, one can assume that all radioactive sources lay within one fixed plane parallel to the detector. Subsequently, CAI can be regarded as an image-to-image mapping and is referred to as planar CAI. Several reconstruction methods for planar CAI have been proposed but yet, no comprehensive quantitative comparison has been carried out to assess their advantages and disadvantages.\\
The main contribution of this paper is a thorough and quantitative comparison of the most commonly used planar coded aperture reconstruction methods. Future research in various fields of nuclear imaging could benefit from this assessment in terms of run time and reconstruction quality. 
An additional contribution of this paper is a data set of high-resolution coded aperture images from three different hot-rod phantoms, acquired with an experimental $\gamma$-camera. Unlike before, this paper shows on experimental data that a Deep Learning approach is capable of delivering better reconstructions than state-of-the-art methods and proved worthy of further investigation. Furthermore, to the best of our knowledge, this paper includes the first re-implementation and application on experimental data of the convolution-based Maximum Likelihood Expectation Maximization algorithm proposed by \cite{Mu2006}.\\
The structure of this paper is as follows: First, section~\ref{sec:state-of-the-art} gives an overview of the mathematical description of CAI and the most commonly used reconstruction methods. Then, section~\ref{sec:mat_and_meths} describes the image acquisition and implementation of both \textit{analytical} and \textit{data-driven reconstruction methods}. Section~\ref{sec:results} presents the reconstruction results in terms of reconstruction quality and run time, which are discussed in section~\ref{sec:discussion}. Finally, section~\ref{sec:conclusion_and_outlook} concludes this paper's main findings and provides an outlook on possible future research. 

\begin{figure}[btp]
\centering 
\includegraphics[width=.9\textwidth]{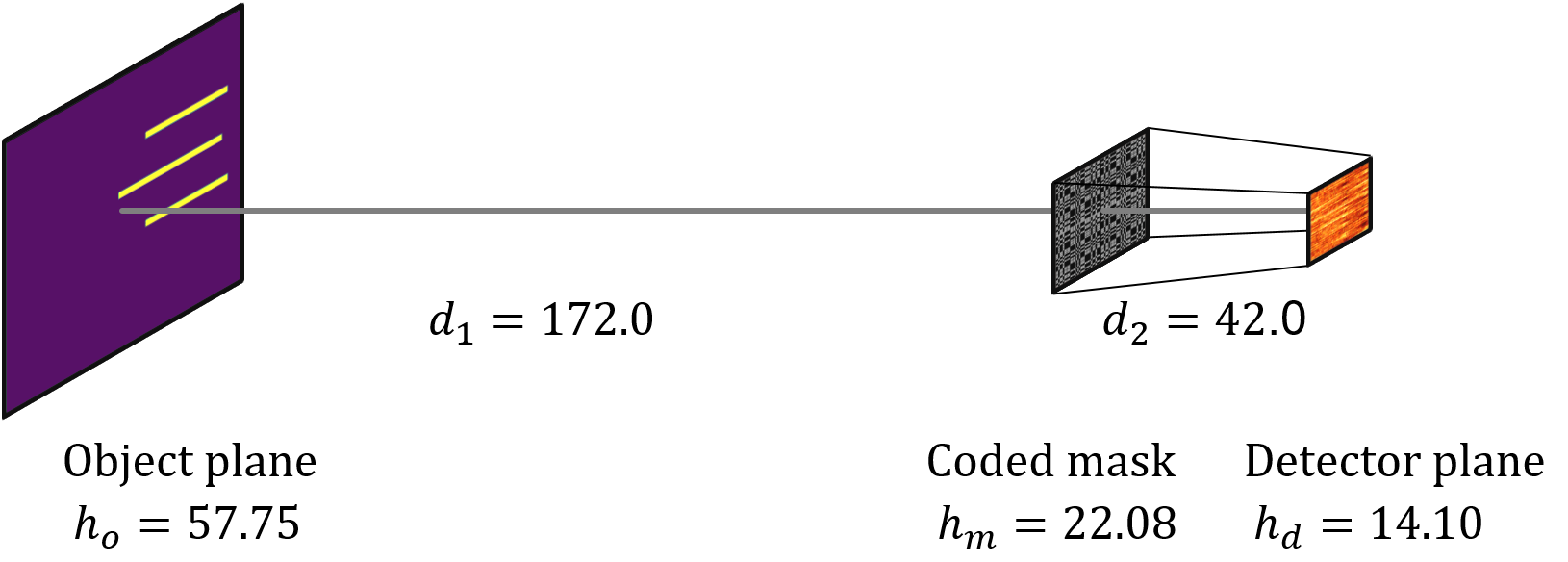}
\caption{\label{fig:y_camera_in_scale} A sketch of the experimental $\gamma$-camera set-up, with all values given in millimeter. Planar Coded Aperture Imaging consists of three parallel planes: The object plane (left), the coded mask (middle), which encodes the object image and the position-sensitive detector (right). The gray line represent the central line. The coded mask and the detector form the $\gamma$-camera itself.}
\end{figure}

\section{State of the art}
\label{sec:state-of-the-art}

\subsection{Mathematical description}
\label{sec:math_description}
Let $\Phi$ be a two-dimensional radioactive source distribution on a plane parallel to the detector plane in a constant and defined distance. The detector image of each point source of $\Phi$ results in a shifted shadow of the coded mask pattern on the detector. As any distribution of sources can be described by a linear combination of point sources, planar CAI can be described as the convolution of an object image $f(x, y)$ with the single coded mask shadow $h(x, y)$ as long as the distance object - mask can be considered to be much larger than the diameter of the mask and as long as scatter at the mask can be neglected \cite{dicke1968scatter}. The resulting detector image is referred to as $p(x, y)$. In reality, this process is not deterministic, but especially in imaging of weakly radiating sources bound to the random Poisson process $\mathfrak{P}$. Additional detector noise $n(x, y)$ must be taken into account, resulting in the following system equation: 

\begin{equation}
\label{eq:cai_as_convolution}
    p(x, y) = \mathfrak{P}\lbrace f(x, y) \ast h(x, y)\rbrace + n(x, y)
\end{equation}

\noindent Multiple families of coded aperture patterns can be found in literature (see \cite{Accorsi2001a} for an overview), but most commonly used are Modified Uniformly Redundant Arrays (MURA), based on the mathematical concept of quadratic residues \cite{Fenimore1978, Gottesman1989}. The pattern can be expressed by a binary matrix, where a “1” denotes a pinhole and “0” opaque material:

\begin{equation}
\label{eq:mura_mask}
    \begin{split}
        H_{ij} = 
        \begin{cases} 
              1 & \text{if i+j=0,} \\
              1 & \text{if $j=0, i \neq 0$,} \\
              1 & \text{if $C_i C_j = 1$},\\
              0 & \text{otherwise}\\
           \end{cases}
       \qquad 
       \text{with} 
       \qquad 
        C_k = 
        \begin{cases} 
            1 & \text{if $k$ is a quadratic residue modulo $L$,} \\
            -1 & \text{otherwise}\\
        \end{cases}
    \end{split}   
\end{equation}

\noindent and zero-based indices $i$ and $j$ referring to rows and columns. L is called the rank and must be both prime and of the following form: $L=4m+1$ with $m$ as natural number. MURA masks reach a fraction of opaque to transparent cells of approximately 50\% \cite{Gottesman1989}. To yield a self-supporting mask, rows and columns of zeros can be inserted between all columns and rows. This so-called no-two-holes-touching (NTHT) mask is easier to fabricate but has the disadvantage of reducing the transparent fraction of this mask to around 12.5\%. The MURA pattern used in this paper can be seen in figure~\ref{fig:triple_mask}.
Commonly, a mosaic of four identical basic patterns is used to assure a complete mask shadow on the detector, while maintaining a wide field-of-view (FoV). The shadow is complete in the sense of circular convolution, which enables decoding only using the decoding pattern of the basic pattern and the central part of the detector \cite{Fenimore1978}. With evolving computing power, a decoding using the complete mosaicked pattern and the whole detector image does not pose a problem anymore and reduces background noise \cite{Mu2006}.\\
It has been proposed to combine two single captures into a dual image acquisition, where one image is taken with the mask and a second image with the anti-mask to remove near-field artifacts \cite{Accorsi2001a,Mu2006,Vassilieva2002}. The anti-mask is the inverse coded aperture mask, which has pinholes where the mask is opaque and vice versa. Certain MURA patterns inhibit the anti-mask in themselves and can be achieved by rotating the mask by 90\degree. This makes dual acquisition easy to implement as no additional mask needs to be manufactured or installed. The reconstructions of the dual images are referred to as \textit{dual image reconstructions}.

\subsection{Reconstruction methods}
\label{sec:reconstruction_methods}
Mainly four methods for real-time capable image reconstruction have been proposed within the last few decades. MURA Decoding (also called inverse filtering, or cross-correlation analysis): \cite{Accorsi2001a, Schellingerhout2002, Garibaldi2005, Accorsi2007, Accorsi2008, Fujii2012, Gmar2011, Vassilieva2002, Kaissas2015, Kaissas2017, Russo2020}, Wiener Filtering \cite{Haboub2014, Kulow2020}, convolutional Maximum Likelihood Expectation Maximization (MLEM) reconstruction \cite{Mu2006, Mu2016} and data-driven Deep Learning approaches \cite{Zhang2019, Zhang2020, Kulow2020}. Although the standard MLEM algorithm has also been investigated \cite{Martineau2010, Jeong2020}, it is not considered in this paper, because its reconstruction time of several hours makes it impractical for the application in mobile systems.

\subsubsection{MURA Decoding}
\label{sec:mura_decoding}
For each MURA mask pattern an inverse decoding pattern $G$ can be determined \cite{Fenimore1978}. Correlation of the decoding and encoding pattern results in a $\delta$-distribution and can be regarded as an inverse filter \cite{beyerer2015machine}. The decoding pattern $G$ in relation to the encoding pattern $H$ equals

\begin{equation}
\label{eq:mura_decoding}
    G_{ij} = 
    \begin{cases} 
          1 & \text{if } i+j=0, \\
          1 & \text{if } H_{ij}=1, i+j \neq 0, \\
          -1 & \text{if } H_{ij}=0, i+j \neq 0. \\
       \end{cases}
\end{equation}

\noindent In practical terms that means all 0 change to -1 and all 1 stay 1 except for the central pixel if a 2\,$\times$\,2 mosaicked mask is chosen \cite{Cieslak2016}. Convolving the given detector image with the decoding pattern and central cropping yields the reconstructed object image.

\subsubsection{Wiener Filter}
\label{sec:wiener_filter}
The Wiener Filter executes reconstruction in an optimal balance between inverse filtering and noise smoothing with regards to the mean squared error. The underlying assumptions are a linear system and Gaussian additive white noise \cite{beyerer2015machine}. It can be regarded as a weighted cross correlation with the mask pattern and has the design parameter signal-to-noise-ratio (SNR). In the Fourier domain with $\nu$ as frequency vector, the Wiener Filter reads as follows:

\begin{equation}
\label{eq:wiener_filter}
    W(\nu)=  \frac{1}{H(\nu)} \frac{|H(\nu)|^2}{|H(\nu)|^2+\frac{1}{\text{SNR}(\nu)})}
\end{equation}

\noindent Unlike MURA Decoding, the reconstruction turns into a multiplication in the Fourier space and thus offers fast reconstruction. However, both methods ignore the random nature of photon emission and detection in nuclear imaging \cite{Mu2006}. 
%The Wiener Filter is rarely used since the proposal of MURA masks and its reconstruction method, but is used in neighboring inverse problems \cite{Wang2011, Martineau2010, Haboub2014} and \cite{Kulow2020} use the Wiener Filter with the assumption of a perfect measurement, where the SNR term vanishes. 

\subsubsection{Maximum Likelihood Expectation Maximization}
\label{sec:mlem}
The Maximum Likelihood Expectation Maximization (MLEM) algorithm is an iterative algorithm that estimates the object image with the highest likelihood for the measured detector image. The algorithm is derived from the assumption that the photon counts follow a Poisson process \cite{Buzug2008}. Originally, probabilistic system knowledge in the form of a transfer matrix $A$ must be provided, where its entries $a_{ij}$ denote the probability that a photon emitted from object pixel $j$ is detected in image pixel $i$. For high-resolution imaging, $A$ would contain billions of entries, rendering the standard MLEM computationally expensive. To overcome this major drawback, Mu \& Hoang \cite{Mu2006} proposed a convolutional-based version of the MLEM algorithm: Instead of using the system matrix $A$ for forward- and back-projection, the mask pattern and the convolution operator is used. %Therefore, the measured detector image $p$ needs to be compensated for angular and collimation first. 

\begin{equation}
\label{eq:mlem}
    \hat{f}^{k+1}(x,y) = \hat{f}^{k}(x,y) \odot \left[\frac{p(x,y)}{ \hat{f}^{k}(x,y) \ast h(x,y)} \otimes h(x,y)\right]
\end{equation}

\noindent with $\odot$ and $\otimes$ denoting element-wise multiplication and the correlation operator. The algorithm consists of two major steps: forward-projection and back-projection. The convolution in the denominator represents the forward-projection step using current estimate $\hat{f}^k$. Dividing the detector image $p$ by the current forward-projection yields the relative difference. The correlation of the relative difference with the mask pattern $h$ represents the back-projection, which is multiplied element-wise with the current estimate $\hat{f}^k$ to obtain an updated estimation $\hat{f}^{k+1}$.

\subsubsection{Convolutional Encoder-Decoder}
\label{sec:cnn}
A Convolutional Encoder-Decoder (CED) is a widely used architecture for Convolutional Neural Networks (CNN). A CED consists of trainable parameters in the form of multiple layers of convolutional kernels that successively transform an input into an output image. The transformation is not derived from a mathematical description of the imaging system. Instead, by providing a sufficient amount of training pairs an approximate transformation from the input domain into the output domain is deduced. Recent advances in Deep Learning in the field of image reconstruction \cite{Tao2018, Haggstrom2019, Belthangady2019, Zhu2018} underline the potential of CEDs for CAI reconstruction. 
By compressing the input image via strided convolution into a latent space representation from which the output image is generated by progressive up-sampling, CEDs effectively regularize the ill-posed inverse mapping by implicit denoising \cite{Belthangady2019}. Thus, CNNs and CEDs in particular, have been deployed successfully on direct reconstruction of positron emission tomography scans from sonograms \cite{Haggstrom2019}, Radon inversion \cite{He2020}, metal artifact reduction in computer tomography \cite{zhang2018convolutional} or image deblurring \cite{Tao2018}, to mention a few examples.\\ 
First experiments have been conducted on the application of CNNs to CAI reconstruction, but they either were on purely simulated and low-resolution images \cite{Zhang2019, Zhang2020} or compared only visually to other methods \cite{Kulow2020}. In the following, the above described reconstruction methods are referred to as \textit{analytical reconstruction methods}, while trained CEDs are denoted by \textit{data-driven reconstruction methods}.

\section{Material \& methods}
\label{sec:mat_and_meths}

\subsection{Experimental measurements}
\label{sec:phantom_measurements}
\begin{figure}[hbpt]
    \centering
    \begin{subfigure}[t]{0.26\textwidth}
        \centering
        % trim=left bottom right top, clip
        \includegraphics[width=\textwidth, trim=7 40 20 30, clip]{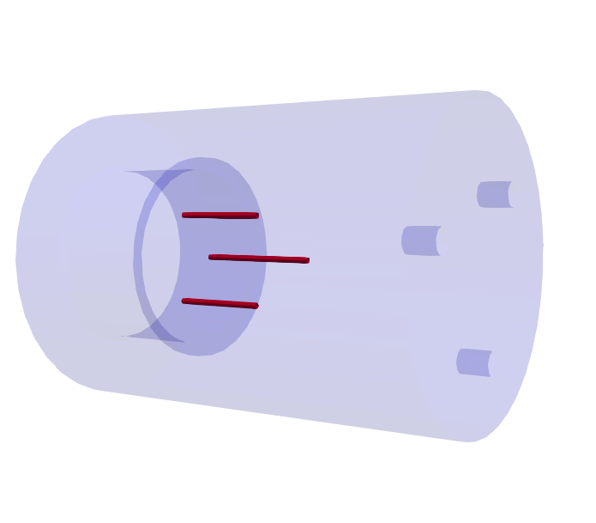}
        \caption{}
        \label{fig:sr_phantom}
    \end{subfigure}
    \hspace{0.0\textwidth}
    \begin{subfigure}[t]{0.26\textwidth}
        \centering
        \includegraphics[width=\textwidth, trim=7 40 20 30, clip]{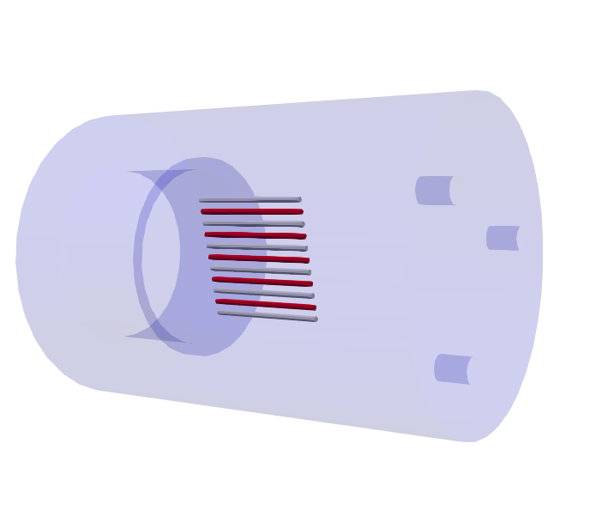}
        \caption{}
        \label{fig:lr_phantom}
    \end{subfigure}
    \hspace{0.0\textwidth}
    \begin{subfigure}[t]{0.26\textwidth}
        \centering
        \includegraphics[width=\textwidth, trim=7 40 20 30, clip]{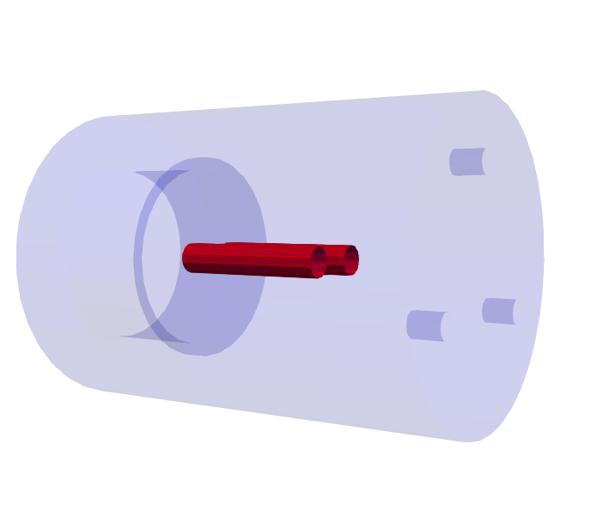}
        \caption{}
        \label{fig:c_phantom}
    \end{subfigure}
    \caption{\label{fig:triple_phantom}Three hot-rod phantoms: (\subref{fig:sr_phantom}) spatial resolution phantom (SRP), (\subref{fig:lr_phantom}) linear resolution phantom (LRP), (\subref{fig:c_phantom}) contrast phantom (CP). Marked in red are the tubes filled with $^{99\text{m}}\text{Tc}$. The indentations on both sides were used to clamp the phantoms.}
\end{figure}

Three different phantoms were designed and manufactured: A spatial resolution phantom (SRP), a linear resolution phantom (LRP) and a contrast phantom (CP). All three phantoms have the basic form of a cylinder with a height of 80\,mm and 50\,mm in diameter, where tubes along the vertical axis were filled with $^{99\text{m}}\text{Tc}$, the most widely used radionuclide in nuclear medicine \cite{Peterson2011}. The total activity was 83\,MBq, 50\,MBq and 75\,MBq for SRP, LRP and CP respectively. Phantoms were designed with a computer-aided-engineering software and afterwards milled out of Lucite. A depiction can be seen in figure~\ref{fig:triple_phantom}. \\
The SRP has three tubes with a diameter of 1.1\,mm, two of which are 15\,mm and one 20\,mm long. The tubes are arranged in parallel at three different positions. The LRP has eleven tubes in a straight line in radial direction. All tubes are 20\,mm long and have a diameter of 1.1\,mm, even though only every second tube was filled with $^{99\text{m}}\text{Tc}$. The CP consists of two larger tubes, which are 25\,mm long and 5\,mm in diameter. They are centered around the vertical axis.\\
The experimental $\gamma$-camera for capturing the phantoms has already been described in \cite{Rozhkov2020} and only the main characteristics are repeated here and are depicted in figure~\ref{fig:y_camera_in_scale}: A 2\,$\times$\,2 mosaicked, 1\,mm thick Tungsten NTHT MURA mask of rank 31 with pinholes of 0.34\,mm in diameter was placed 42\,mm in front of a 2\,mm thick 256\,$\times$\,256 pixelated CdTe semiconductor detector coupled to a Timepix\textsuperscript{\textcopyright} readout circuit. The virtual object plane is 172\,mm in front of the mask plane, resulting in a FoV of 57.75\,$\times$\,57.75\,mm. The mask is anti-symmetrical and can be rotated by 90\degree to form the anti-mask.
Each phantom was placed on a rotational table in front of the camera. The detector was exposed for 2\,min and afterwards the phantom was rotated along its vertical axis by 3\degree. One complete rotation per hot-rod phantom resulted in 120 images per phantom. Each phantom was captured once with mask and once with anti-mask and thus a total amount of 720 images were acquired. Detected photons outside the energy window of 10\,keV centered at the photon peak of 140\,keV for $^{99\text{m}}\text{Tc}$ were discarded.\\
\noindent To remove erroneous pixels, a global thresholding followed by pixel replacement was carried out. The average image of all 720 captures, was thresholded according to the 2\textsuperscript{nd} and the 98\textsuperscript{th} percentile. Pixels outside this range were considered outliers and replaced by the median of their 3\,$\times$\,3 neighborhood.

\subsection{Generation of ground truth data}
\label{sec:generation_of_gt}
In order to compare reconstruction methods, the perfect reconstruction (the so-called ground truth) is essential.
Therefore, the phantom models were loaded into a visualization program (Paraview Version 5.9.1 \cite{ahrens2005paraview}), where all surfaces except for the rods were set transparent. The point of view, i.e. camera position, focal point and view angle were set according to the experimental set-up. The background and the hot-rod color was set to black and white and the camera resolution to 256\,$\times$\,256. By using automated animation, the phantom was rotated along the horizontal axis and at each position the rendered image was stored as image file. To obtain fully binarized images and remove shadows, those images were cleaned by applying a simple thresholding. The results are binary images, which are 1 where $\gamma$-sources are expected and 0 elsewhere. The images were used as ground truth data for both the single and dual image reconstruction.

\subsection{Analytical reconstruction methods}
\label{sec:analytical_reconstruction_methods}
For all reconstruction methods, not just the central part, but as suggested by \cite{Mu2006} the complete detector image and mask pattern was used for reconstruction. This allowed usage of the entire detector and FoV.
The reconstructions of each method were finally normalized to the range of [0, 1] to assure a fair comparison. Even though all reconstruction methods were intended for dual image reconstruction, the application to the single images was also analyzed.\\
The inference time per method was measured by averaging the elapsed time for the reconstruction of one image over 1,000 runs on the CPU of a computer with a 6-kernel Intel Core i7 processor (2.6\,GHz) with six kernels and 16\,GB of RAM. TensorFlow allows the use of a GPU to speed up matrix and vector calculations, thus the inference time with GPU support was also evaluated with a Nvidia GeForce RTX 2070 and 8\,GB dedicated RAM. All processing and implementation was carried out in TensorFlow 2.7 \cite{abadi2016tensorflow} and Python 3.9.7. 

\noindent\textbf{MURA Decoding:} Two different decoding patterns were constructed. A two-holes-touching (THT) and a not-two-holes-touching (NTHT) pattern based on the basic MURA pattern of rank 31 from eq.~\ref{eq:mura_mask}. Both have been used as base pattern for reconstruction in literature \cite{Haboub2014, Accorsi2008}. 
After converting the mosaicked pattern into the decoding pattern by eq.~\ref{eq:mura_decoding}, it was resized to 512\,$\times$\,512 pixels by nearest-neighbor interpolation to assure a binary final decoding pattern. For the reconstruction of single acquisition images, the decoding pattern can directly be convolved with the detector image to obtain the reconstructed image. 
For the reconstruction using the dual image, both images are reconstructed separately and summed up afterwards. However, due to the linear operation and the relationship from mask and anti-mask, it is equivalent to applying it once to the subtraction of both single images \cite{Vassilieva2002}. 

\begin{figure}[bpt]
    \centering
    \includegraphics[width=0.9\textwidth, trim=0 0 0 0, clip]{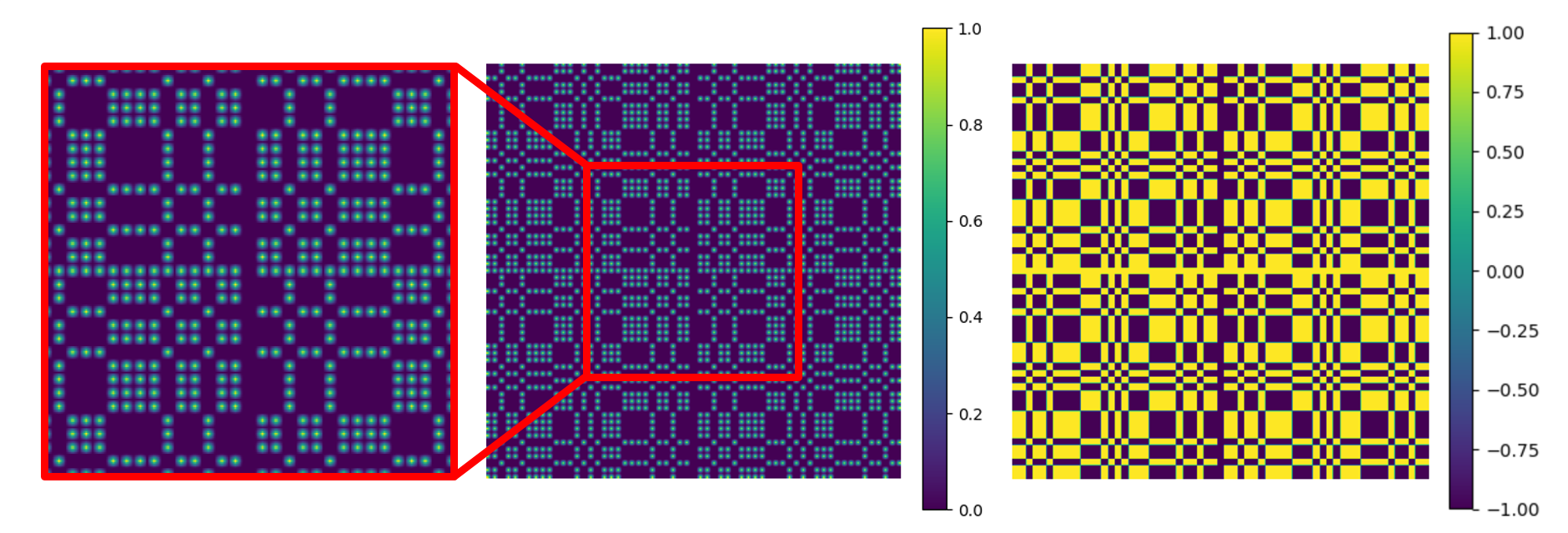}
    \caption{\label{fig:triple_mask} From left to right: The centralized basic MURA pattern (rank 31), its 2\,$\times$\,2 arrangement with bilinear resizing for simulated forward-projection and the according THT decoding pattern for MURA Decoding.}
\end{figure}

\noindent\textbf{Wiener Filter:} The Wiener Filter was calculated according to eq.~\ref{eq:wiener_filter} with various SNR. Analogously to MURA Decoding, two types of mask pattern were used as base pattern for the Wiener Filter: The THT and the NTHT version. In order to avoid aliasing, 
%exploit the complete detector image, 
the detector image was periodically padded to 512\,$\times$\,512 pixels and transformed into Fourier domain, similar to \cite{Haboub2014}. After element-wise multiplication with the Wiener Filter, the result was transformed back into the spatial domain, where the central 256\,$\times$\,256 pixels contained the final reconstructed image.
A line search was carried out and $\text{SNR} = 10^{-7}$ was chosen and used for all succeeding analyses. Dual image reconstruction was carried out analogously to MURA Decoding.

\noindent\textbf{MLEM:} The algorithm according to eq.~\ref{eq:mlem} was implemented with an additional small $\varepsilon$ of $10^{-7}$ added to the denominator to avoid zero divisions. Both the THT and the NTHT version of the MURA pattern were tested for the MLEM algorithm. Instead of applying adaptive stopping criteria, a fixed number of 25 iterations was chosen heuristically, which confirmed to be a good compromise between reconstruction quality and noise amplification. The initial guess was set to a constant image of 0.5. 
Unlike MURA Decoding and the Wiener Filter, the MLEM algorithm is a non-linear operation. Thus, both single acquisition images were reconstructed independently from each other, with their corresponding mask and anti-mask. Afterwards, the final reconstructed image is obtained by averaging both reconstructions, as suggested by \cite{Mu2006}.

\subsection{Data-driven reconstruction methods}
\label{sec:Data-driven reconstruction methods}
\subsubsection{Generation of training data}
\label{sec:generation_of_training_data}
What separates the Machine Learning approach from analytical methods is the need for a large quantity of image pairs for training the network, consisting of an input image (the detector image) and a target image (the reconstructed image). A common way to solve the data problem is by using simulated training data \cite{Haggstrom2019, Zhang2020, Kulow2020}:
In this paper, two different sets of training data were generated: One based on the large collection of natural photographs called \textit{ImageNet} \cite{Russakovsky2015} and one based on horizontal lines on dark background:
For the first set of training data, 50,000 images from the ImageNet’s validation set were cropped to 256\,$\times$\,256 pixels and normalized to the range of 0 to 1. Those images posed as object images. For the second set of training data, 50,000 images with black background and different number of horizontal lines with varying thickness were created using the openCV library version 3.4.2 \cite{bradski2000opencv}. The number of lines varied between 2 and 6, with a height and width between 4 and 24 pixels and between 60 and 122 pixels, respectively. The center points were chosen randomly over the whole image. Each line has a uniform intensity between 0 and 1 and the intensities of overlapping lines were added, to ensure a non-binary intensity distribution. The final image was normalized.

\begin{figure}[bpt]
    \centering
    \begin{subfigure}[t]{0.26\textwidth}
        \centering
        % trim=left bottom right top, clip
        \includegraphics[width=\textwidth, trim=5 2 5 2,clip]{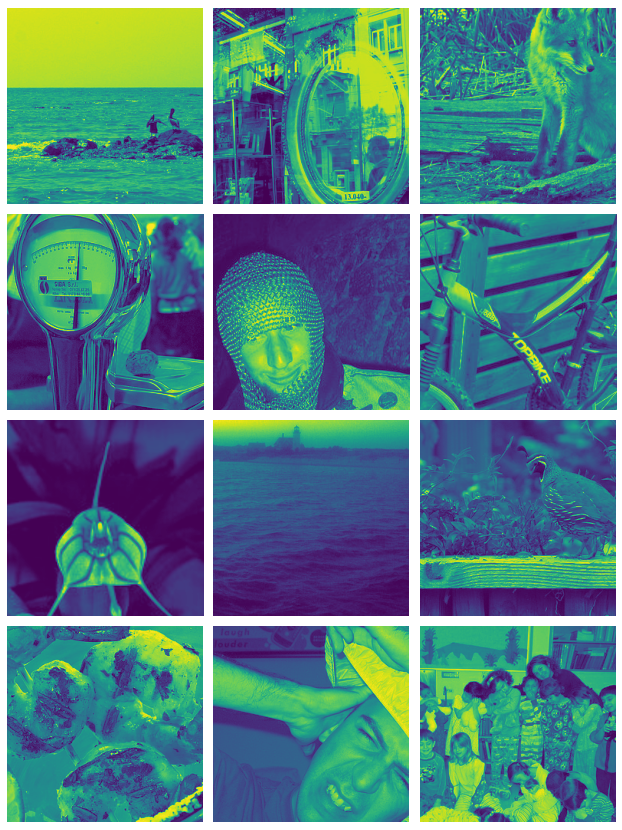}
        \caption{}
        \label{fig:Imagenet_output_set}
    \end{subfigure}
    \hspace{0.05\textwidth}
    \begin{subfigure}[t]{0.26\textwidth}
        \centering
        \includegraphics[width=\textwidth, trim=5 2 5 2,clip]{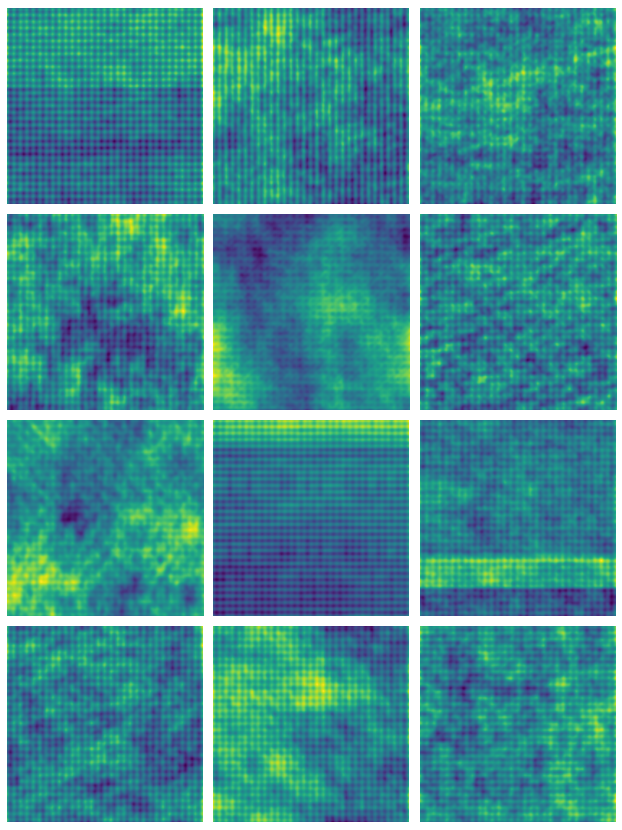}
        \caption{}
        \label{fig:Imagenet_input_set}
    \end{subfigure}
    \hspace{0.05\textwidth}
    \begin{subfigure}[t]{0.26\textwidth}
        \centering
        \includegraphics[width=\textwidth, trim=5 2 5 2,clip]{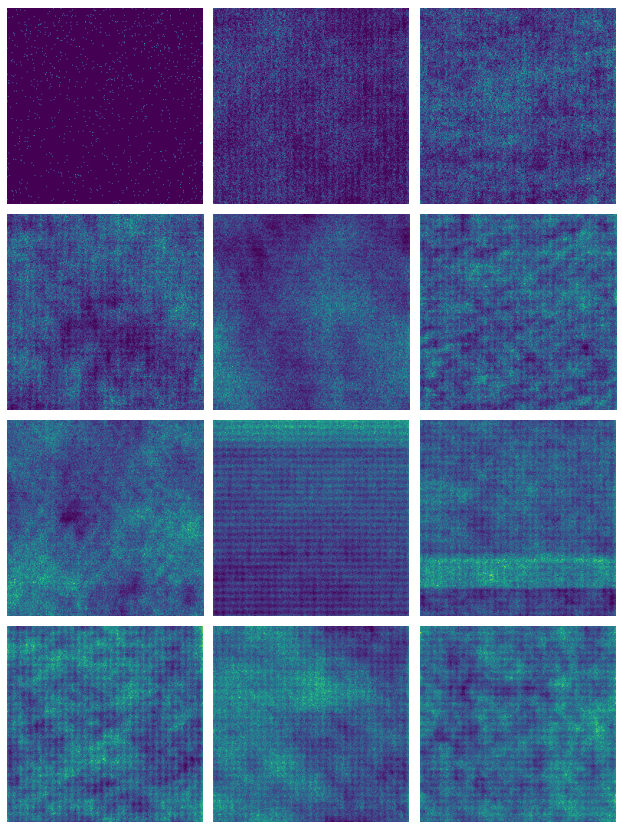}
        \caption{}
        \label{fig:Imagenet_input_set_blurry}
    \end{subfigure}
    \caption{\label{fig:triple_Imagenet} (\subref{fig:Imagenet_output_set}) shows exemplary target images from the ImageNet based training set, (\subref{fig:Imagenet_input_set}) their forward-projection and (\subref{fig:Imagenet_input_set_blurry}) the final training input image after Poisson randomization with an increasing photon count from top left to bottom right.}
\end{figure}

\begin{figure}[bpt]
\centering
    \begin{subfigure}[t]{0.26\textwidth}
        \centering
        % trim=left bottom right top, clip
        \includegraphics[width=\textwidth, trim=5 2 5 2,clip]{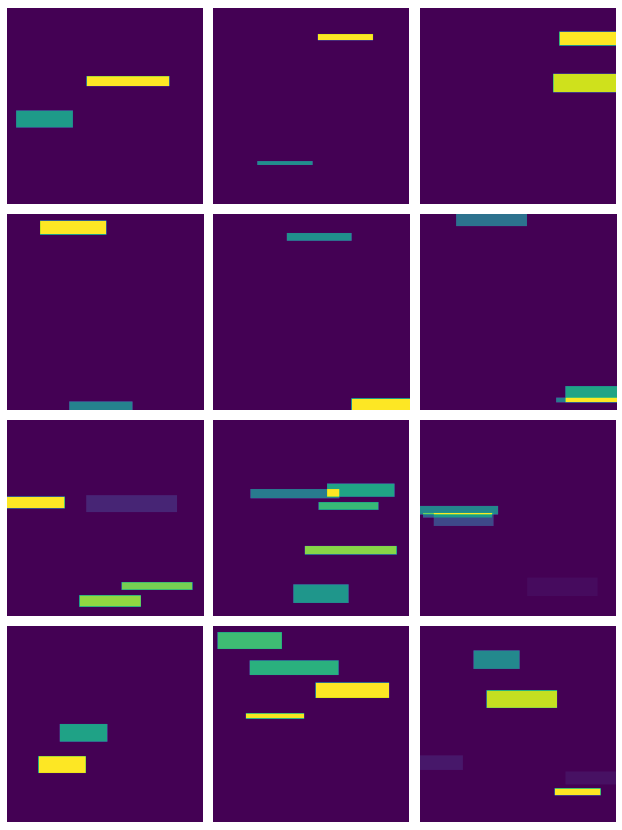}
        \caption{}
        \label{fig:RozhkovLines_output_set}
    \end{subfigure}
    \hspace{0.05\textwidth}
    %\hfill
    \begin{subfigure}[t]{0.26\textwidth}
        \centering
        \includegraphics[width=\textwidth, trim=5 2 5 2,clip]{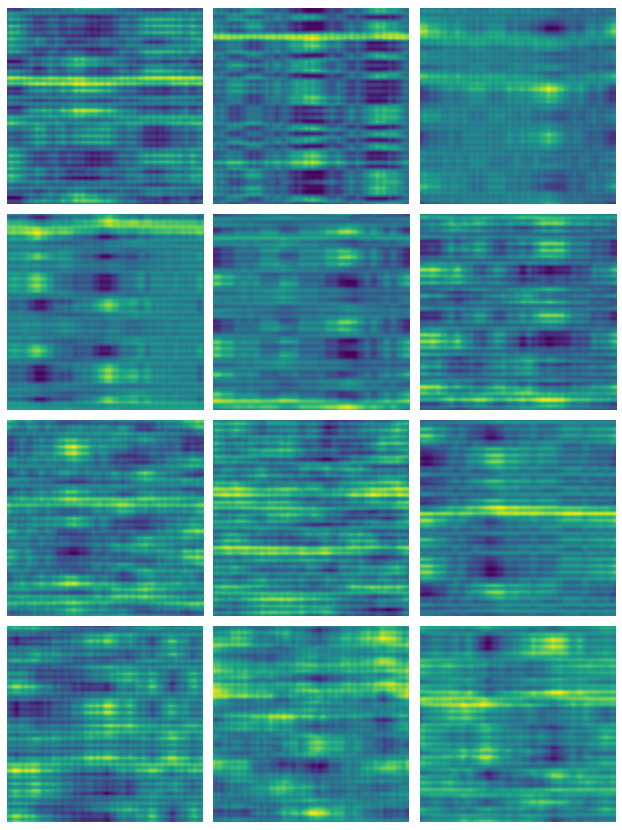}
        \caption{}
        \label{fig:RozhkovLines_input_set}
    \end{subfigure}
    \hspace{0.05\textwidth}
    \begin{subfigure}[t]{0.26\textwidth}
         \centering
         \includegraphics[width=\textwidth, trim=5 2 5 2,clip]{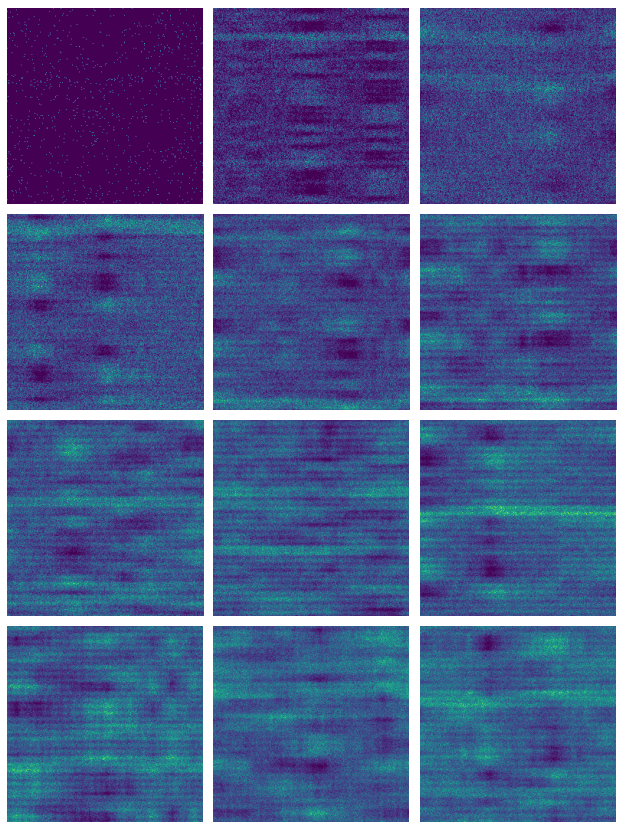}
         \caption{.}
         \label{fig:RozhkovLines_input_set_blurry}
    \end{subfigure}
    \caption{\label{fig:triple_Rozhkov} (\subref{fig:RozhkovLines_output_set}) shows exemplary target images from the horizontal line training set, (\subref{fig:RozhkovLines_input_set}) their forward-projection and (\subref{fig:RozhkovLines_input_set_blurry}) the final training input image after Poisson randomization with an increasing photon count from top left to bottom right.}
\end{figure}

\noindent The forward-projection, which results in the simulated detector image, was carried out according to eq.~\ref{eq:cai_as_convolution}. The NTHT MURA mask pattern of the experimental $\gamma$-camera was generated. Similar to \cite{Kulow2020} the binary pattern needed to be rescaled with respect to the object-mask and mask-detector distances where two conditions needed to be fulfilled: First, the final mask pattern is required to be four times the size of the detector in pixels, so that the basic MURA pattern covers the entire detector. Second, the illuminated circle produced by one pinhole must match the magnified pinhole size: The illuminated circle in pixel was calculated according to the pinhole diameter (0.34\,mm), the system's magnification factor (1.244) and the detector pixel size (0.0551\,mm/pixel). 
Bilinear resizing of the NTHT MURA mask to 512\,$\times$\,512 pixels fulfilled both conditions and yielded the final mask array. The obtained matrix can be regarded as the complete point spread function (PSF) of the $\gamma$-camera and is depicted in the middle of figure~\ref{fig:triple_mask}. By convolving an object image with this complete PSF and afterwards cropping the central 256\,$\times$\,256 pixels, one yields the detector image of the given object image. The additive noise $n(x, y)$ was set to be zero, assuming a ultra-low noise detector with single photon detection.\\
Different levels of Poisson noise were simulated: The sum of all pixel values, corresponding to the amount of photons the detector has detected, was scaled to reach a random number between the range of 1,000 and 1,000,000 photons. This scaled image represents the \textit{expected average photon count per pixel} and formed the basis for a random draw from a Poisson distribution $\mathfrak{P}$. \\
For training the dual image reconstruction CED, the same forward-projection was carried out with the anti-mask as well and concatenated as a second color channel. 
Finally, the simulated detector image was normalized to the range of 0 to 1 for each channel. Exemplary images with their simulated forward-projection with and without Poisson randomization from the ImageNet data set and for the horizontal line data set can be seen in figure~\ref{fig:triple_Imagenet} and figure~\ref{fig:triple_Rozhkov}. The generation of training data was done on-the-fly while training the CEDs.

\subsubsection{Architecture and training}
\label{sec:ced_and_its_training}
\begin{figure}[btp]
     \centering
     % trim=left bottom right top, clip
     \includegraphics[width=0.8\textwidth, trim=0 0 10 0,clip]{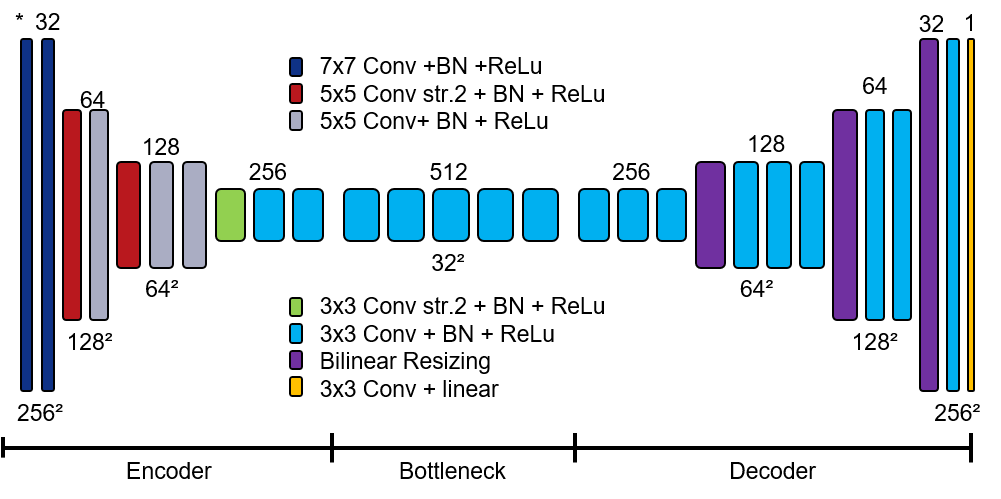}
     \caption{\label{fig:ced_architecture} The architecture of the deployed CED network with the number of feature channels on top and the spatial size at the bottom. The asterisk indicates the input image channel size of 1 or 2 for single and dual image reconstruction respectively.}
\end{figure}

Figure~\ref{fig:ced_architecture} presents the Convolutional Encoder-Decoder (CED) architecture used in this paper. It is based on \cite{Ronneberger2015, Jin2017, He2020} and in particular on \cite{Haggstrom2019}. It was decided to discard the use of skip or additive connections, as there is no similarity on a pixel level between input and output domain. The main building block is a convolution block. It consists of a convolutional layer with different kernel and strides, batch normalization \cite{pmlr-v37-ioffe15} and is concluded by a ReLU activation function \cite{nair2010rectified}. A large kernel size of 7\,$\times$\,7 pixels and 5\,$\times$\,5 pixels in the encoding part ensures a large receptive field, which is especially important in CAI, as information is spread across the entire input image. The input image has either one or two channels for single and dual image reconstruction.
During encoding, the spatial resolution is successively reduced to 32\,$\times$\,32 pixels by strided convolution. Simultaneously, the number of channels increases from 1 or 2 to eventually 512 in the bottleneck. 

\noindent The bottleneck was intended for the actual transformation between detector domain and object domain and thus the majority of trainable parameters were located here.
The decoder takes the encoded representation in the latent space and gradually increases the spatial size, while reducing the channels to reach a final output image with a single channel and 256\,$\times$\,256 pixels. As described in \cite{Odena2017}, using transposed convolutions can lead to checkerboard artifacts in the output image, when performing image regression tasks. Therefore, bilinear resizing followed by unstrided convolutions was employed in this network’s decoder. The kernel size remains 3\,$\times$\,3 for the complete decoder part, to produce high-resolution images. The final layer is a convolution followed by a linear activation function, because of its constant sensitivity. Overall, the CED architecture for single image reconstruction has a total of 16,414,817 trainable parameters. For dual image reconstruction this number increased to 16,416,385. \\
The CED, as well as all other reconstruction methods, were implemented, trained and evaluated in TensorFlow. The network intended for single image reconstruction and trained on the ImageNet data base is referred to as \textit{CED-IN}, the network additionally fine-tuned on the horizontal lines data set as \textit{CED-Lines} and the networks for dual image reconstruction as \textit{DualCED-IN} and as \textit{DualCED-Lines}. In the following both trained CEDs are considered as separate reconstruction methods.\\
The loss function used for training is the mean squared error of output and target image and was optimized by the Adam optimizer \cite{Kingma2015} with its standard parameters. A mini-batch size of 16 was chosen, as a compromise between short training time and a small generalization gap between training and validation data \cite{Goodfellow2016}. The network was trained on 35,000 training images and validated during training on separate 10,000 validation images for 20 epochs. If within three epochs the validation loss had not improved, the learning rate was automatically reduced by a factor of $0.4$.\\ 
Training was carried out for both training sets, whereas the network trained on the ImageNet data set served as pre-trained network for the horizontal line data set. Afterwards, the CEDs with the best performance on the validation set were chosen and applied to the hot-rod phantom data.

\subsection{Contrast-to-noise ratio}
\label{sec:performance_metric}
The contrast-to-noise-ratio (CNR) takes the source visibility and the degradation from noise into consideration and is thus a suitable metric to evaluate the reconstruction quality. The following definition of CNR is employed \cite{Zhang2020}:

\begin{equation}
\label{eq:cnr}
    CNR = \frac{\left|\bar{S} - \bar{B}\right|}{\sigma}\,, 
\end{equation}

\noindent where $\bar{S}$ denotes the mean intensity of the signal, $\bar{B}$ the mean intensity and $\sigma$ the standard deviation of the background. The binary ground truth images from Section~\ref{sec:generation_of_gt} allowed the division of reconstructions into a signal part and a background part.

\section{Results}
\label{sec:results}

\subsection{Hot-rod phantom data}
\label{sec:hot_rod_phantom_data}
\begin{figure}[hbtp]
    \centering
    \begin{subfigure}[b]{0.5\textwidth}
        \raggedleft
         % trim=left bottom right top, clip
        \includegraphics[width=\textwidth, trim=15 10 15 5,clip]{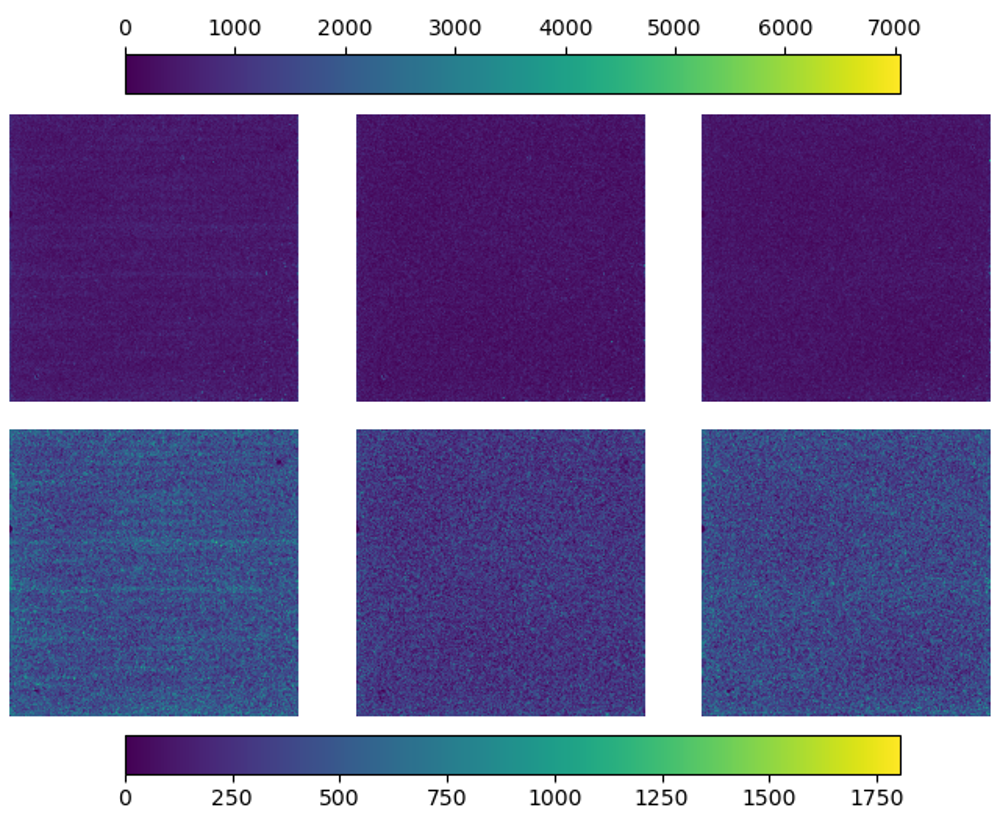}
        \caption{}
        \label{fig:detector_preprocessed}
     \end{subfigure}
     \hspace{-0.0\textwidth}
     \begin{subfigure}[b]{0.4\textwidth}
        \raggedright
        \includegraphics[width=\textwidth, trim=50 10 55 10, clip]{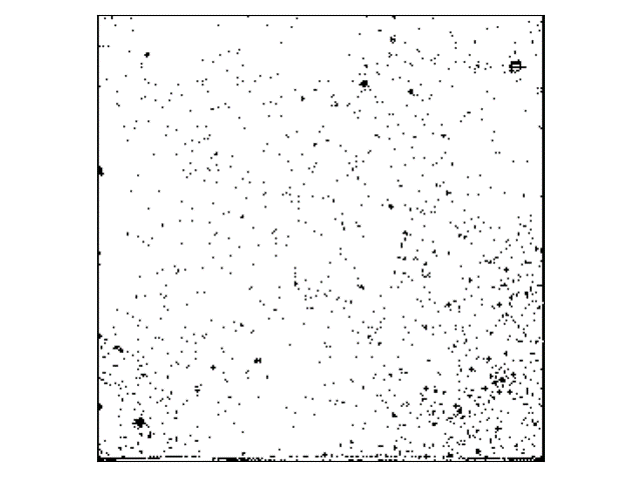}
        \caption{}
        \label{fig:detector_mean_mask_p2p98}
    \end{subfigure}
    \caption{\label{fig:double_phantom_data} (\subref{fig:detector_preprocessed}) shows the first detector images from each of the three captured hot-rod phantoms before (top) and after post-processing. (\subref{fig:detector_mean_mask_p2p98}) shows the binary mask which mark outliers in black according to the 2\textsuperscript{nd} and 98\textsuperscript{th} percentile of the average detector image.}
\end{figure}

The first detector image for each of the three hot-rod phantoms is presented in figure~\ref{fig:detector_preprocessed} before and after post-processing together with the mask used for global outlier replacement (figure~\ref{fig:detector_mean_mask_p2p98}). After capturing the data, pixels with high intensity corrupt the image and structure is barely visible as the top row shows. Replacing outliers by the median value of their neighborhood lead to a higher contrast and vertical stripes becoming visible.\\
The binary mask generated by thresholding the average detector image (figure~\ref{fig:detector_mean_mask_p2p98}) shows multiple clusters of erroneous pixels in addition to pixels along the edge. Despite the post-processing, systematic pixel errors remain and are noticeable as black spots in the top right corner of all detector images of figure~\ref{fig:BMW_overview_single_acq}.

\subsection{Single image reconstruction}
\label{sec:single_image_recon}
\begin{figure}[bpt]
     \centering
     % trim=left bottom right top, clip
     \includegraphics[width=0.9\textwidth, trim=10 10 10 22,clip]{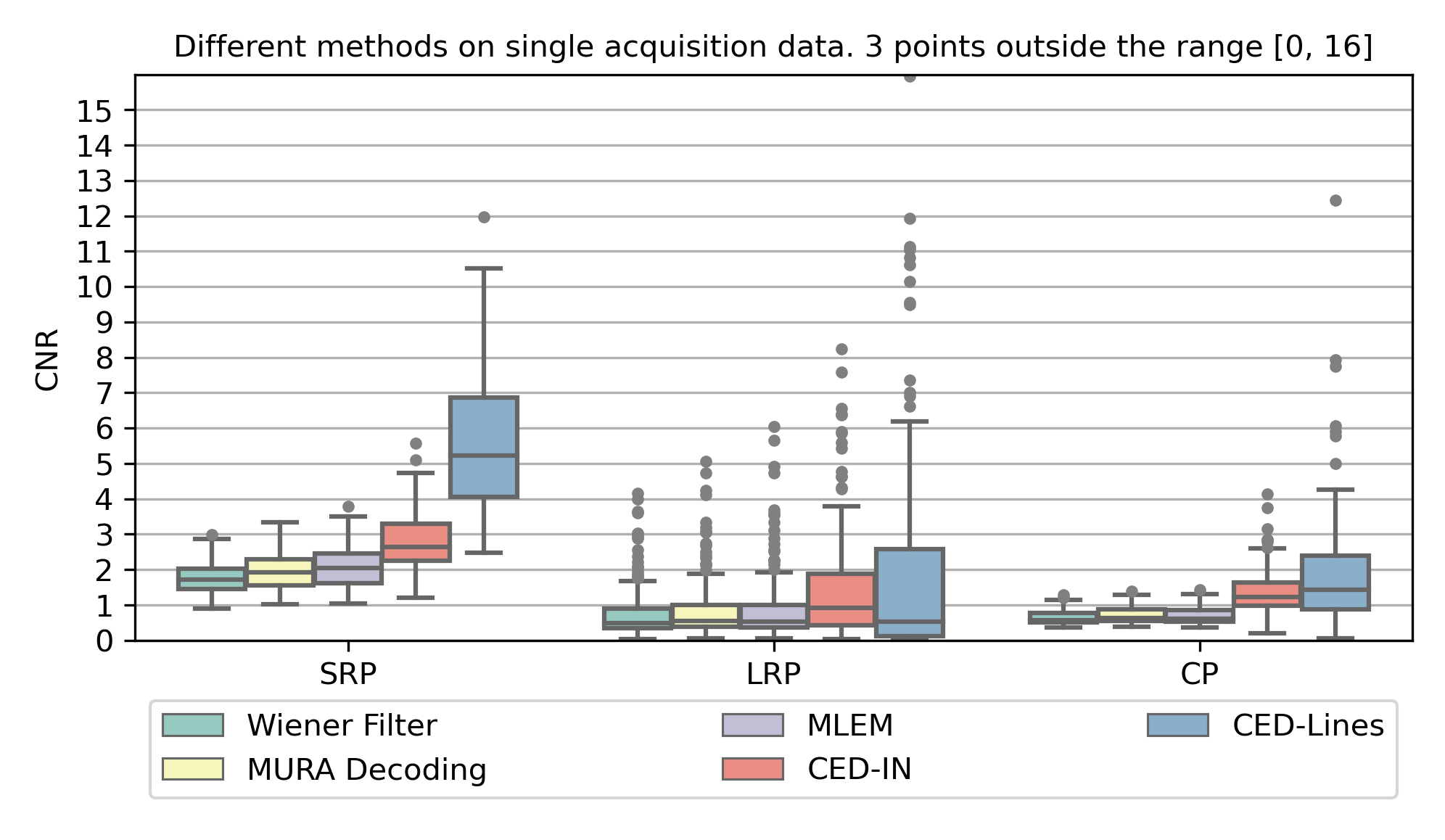}
     \caption{\label{fig:boxplot_single_acq} CNR distribution of the reconstructions of single acquisition images separated by the three hot-rod phantoms (SRP, LRP and CP). Three data points above a CNR of 16 are not shown.}
\end{figure}

\begin{figure}[bpt]
    \centering
     % trim=left bottom right top, clip
    \includegraphics[width=0.9\textwidth, trim=4 2 5 2,clip]{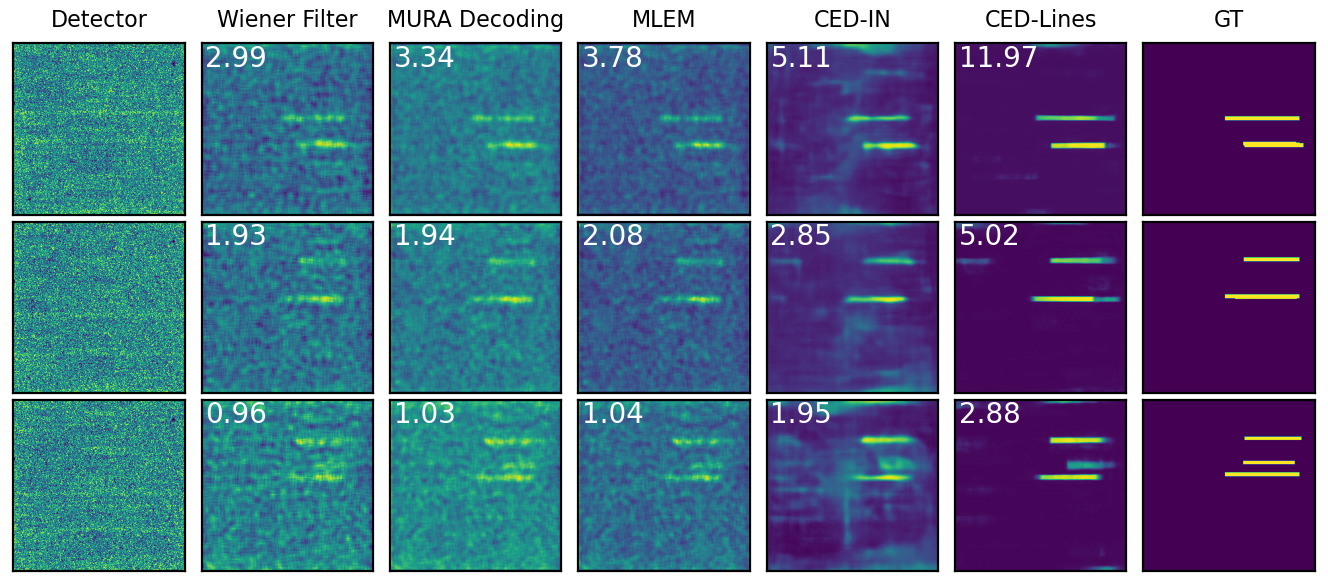}
    \includegraphics[width=0.9\textwidth, trim=4 2 5 2,clip]{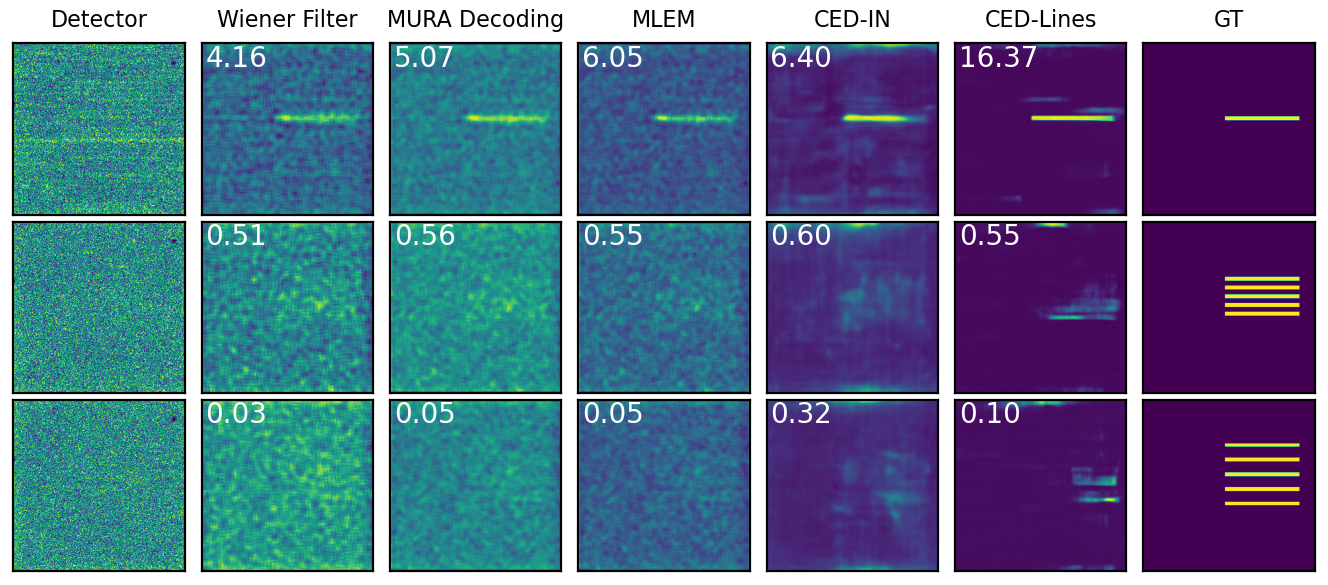}
    \includegraphics[width=0.9\textwidth, trim=4 2 5 2,clip]{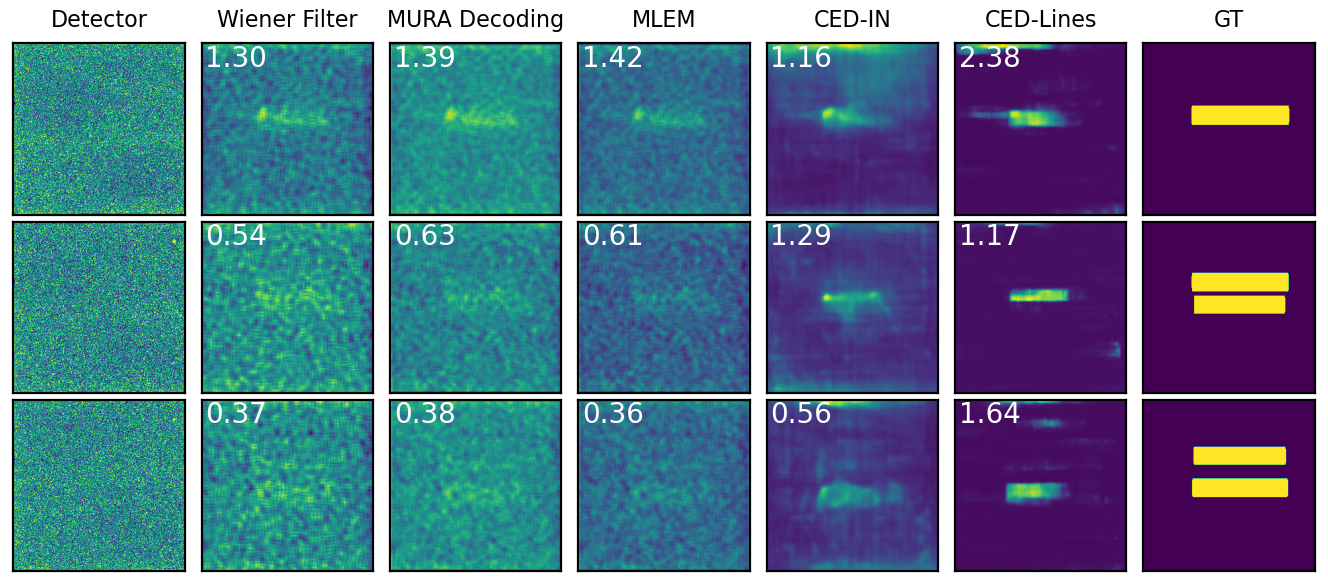}
    \caption{Exemplary reconstructions of single acquisition images from the SRP (top), LRP (middle) and CP (bottom). For each hot-rod phantom from top to bottom row: The best, median and worst reconstructions from MURA Decoding and the corresponding reconstructions from all other methods are displayed. The outer columns represents the detector and ground truth (GT) image respectively. Note, that the histogram of the detector images were equalized for better visualization.}
    \label{fig:BMW_overview_single_acq}
\end{figure}

Training the CED-IN took roughly 11:55\,h and additionally 3:45\,h for the horizontal line data set. Generally, for all analytical reconstruction methods, the THT mask pattern for reconstruction obtained a higher contrast-to-noise ratio (CNR) than its NTHT version and is presented here. The distribution of the 120 CNR values for the five different reconstruction methods are presented as grouped boxplots in figure~\ref{fig:boxplot_single_acq}. The median CNR values for each reconstruction method and phantom is listed on the left side of table~\ref{tab:cnr_single_dual}. \\
In general, it is noticeable, that the SRP is reconstructed with a higher CNR than the other two phantoms: While the median CNR for the SRP varies between 1.72 and 5.22, the median CNR for the CP ranges between 0.57 and 1.42 and is even smaller for the LRP with median CNRs below 1.0.\\
The Wiener Filter and MURA Decoding show similar reconstruction results, with the latter producing a smoother background and hence reaching an overall higher median CNR of for all phantoms. Among the analytical reconstruction methods however, the MLEM algorithm yields the best reconstructions for the SRP and reaches a similar quality for the remaining phantoms. The MLEM algorithm produces reconstructions with a darker and more uniform background, but also slightly weaker tubes. \\
Both data-driven reconstruction methods reconstructed the phantoms with a higher CNR than all analytical reconstruction methods for two of the three phantoms, namely the SRP and CP. While the CED-IN is able to achieve a median CNR of 2.65 for the SRP, the CED-Lines achieves an even higher median of 5.22. Only for the LRP, the CEDs did not outperform the analytical reconstruction methods. The background generated by the data-driven reconstruction methods and especially for the CED-Lines is almost uniformly dark, except for few cloudy artifacts. \\
As figure~\ref{fig:BMW_overview_single_acq} shows, the tubes of the SRP are clearly visible in all reconstructions, but are less visible for the LRP and CP, especially when the tubes are in an upright position. The second tube of the CP is missing from the reconstructions overall. Striking is a thin bright stripe on the top and in few cases also at the bottom of each image in most of the reconstructions, independent from the reconstruction method.\\
The average run times and their standard deviations are presented in table~\ref{tab:run_times}: 25 iterations of MLEM executed on the CPU takes an average of 12,995\,ms, while MURA Decoding takes 288\,ms and the Wiener Filter even less with 67\,ms. The run time of both CEDs are almost identical with 285\,ms and 286\,ms on the CPU and 52\,ms for both CEDs on the GPU. When deploying a GPU, the run times decrease, but are generally associated with a higher standard deviation. 

\begin{table}[hbpt]
\centering
\caption{The median CNR separated in single or dual image reconstruction and hot-rod phantoms. The reconstruction method with the highest CNR is printed in bold. }
\label{tab:cnr_single_dual}
\begin{tabularx}{0.8\textwidth}{lXXXXXX} 
\toprule
\multirow{2.25}{0pt}{\textbf{Reconstruction method}} & \multicolumn{3}{c}{\textbf{single reconstruction}} & \multicolumn{3}{c}{\textbf{dual reconstruction}} \\
\cmidrule(lr){2-4}
\cmidrule(lr){5-7}
& \textbf{SRP}      & \textbf{LRP}      & \textbf{CP}       & \textbf{SRP}      & \textbf{LRP}     & \textbf{CP} \\ \midrule
\textbf{Wiener Filter} & 1.72              & 0.50              & 0.57              & 2.22              & 1.02             & 0.89        \\
\textbf{MURA Decoding} & 1.92              & 0.56              & 0.62              & 2.60              & 1.24             & 1.10        \\
\textbf{MLEM}          & 2.04              & 0.53              & 0.61              & 1.98              & 0.92             & 0.81        \\
\textbf{Dual/CED-IN}   & 2.65              & \textbf{0.92}     & 1.23              & 4.00              & 1.54             & 2.37              \\
\textbf{Dual/CED-Lines}& \textbf{5.22}     & 0.52              & \textbf{1.42}     & \textbf{6.78}     & \textbf{1.71}    & \textbf{2.86} \\
\bottomrule
\end{tabularx}
\end{table}

\begin{table}[hbpt]
\centering
\caption{Average run time of each algorithm in milliseconds on the computer specifications described in the text.}
\label{tab:run_times}
\begin{tabularx}{0.8\textwidth}{lXXXX}
\toprule
\multirow{2.25}{0pt}{\textbf{Reconstruction method}} & \multicolumn{2}{c}{\textbf{single reconstruction}}  & \multicolumn{2}{c}{\textbf{dual reconstruction}}         \\
\cmidrule(lr){2-3}
\cmidrule(lr){4-5}
& \multicolumn{1}{c}{\textbf{CPU}}        & \multicolumn{1}{c}{\textbf{GPU}} & \multicolumn{1}{c}{\textbf{CPU}}          & \multicolumn{1}{c}{\textbf{GPU}} \\
\midrule
\textbf{Wiener Filter}  & \multicolumn{1}{r}{67\,$\pm$\,6}                 & \multicolumn{1}{r}{37\,$\pm$\,44}       & \multicolumn{1}{r}{66\,$\pm$\,6}               & \multicolumn{1}{r}{35\,$\pm$\,35}       \\
\textbf{MURA Decoding}  & \multicolumn{1}{r}{288\,$\pm$\,8}                & \multicolumn{1}{r}{53\,$\pm$\,56}       & \multicolumn{1}{r}{311\,$\pm$\,14}             & \multicolumn{1}{r}{55\,$\pm$\,106}         \\
\textbf{MLEM}           & \multicolumn{1}{r}{12,995\,$\pm$\,345}           & \multicolumn{1}{r}{823\,$\pm$\,12}      & \multicolumn{1}{r}{17,905\,$\pm$\,299}          & \multicolumn{1}{r}{1,097\,$\pm$\,158}    \\
\textbf{Dual/CED-IN}    & \multicolumn{1}{r}{285\,$\pm$\,10}               & \multicolumn{1}{r}{52\,$\pm$\,37}       & \multicolumn{1}{r}{290\,$\pm$\,9}              & \multicolumn{1}{r}{51\,$\pm$\,18}       \\
\textbf{Dual/CED-Lines} & \multicolumn{1}{r}{286\,$\pm$\,9}                & \multicolumn{1}{r}{52\,$\pm$\,17}       & \multicolumn{1}{r}{300\,$\pm$\,27}             & \multicolumn{1}{r}{53\,$\pm$\,37}       \\              
\bottomrule
\end{tabularx}
\end{table}

\subsection{Dual image reconstruction}
\label{sec:dual_image_recon}
\begin{figure}[bpt]
     \centering
     % trim=left bottom right top, clip
     \includegraphics[width=0.9\textwidth, trim=10 10 10 22,clip]{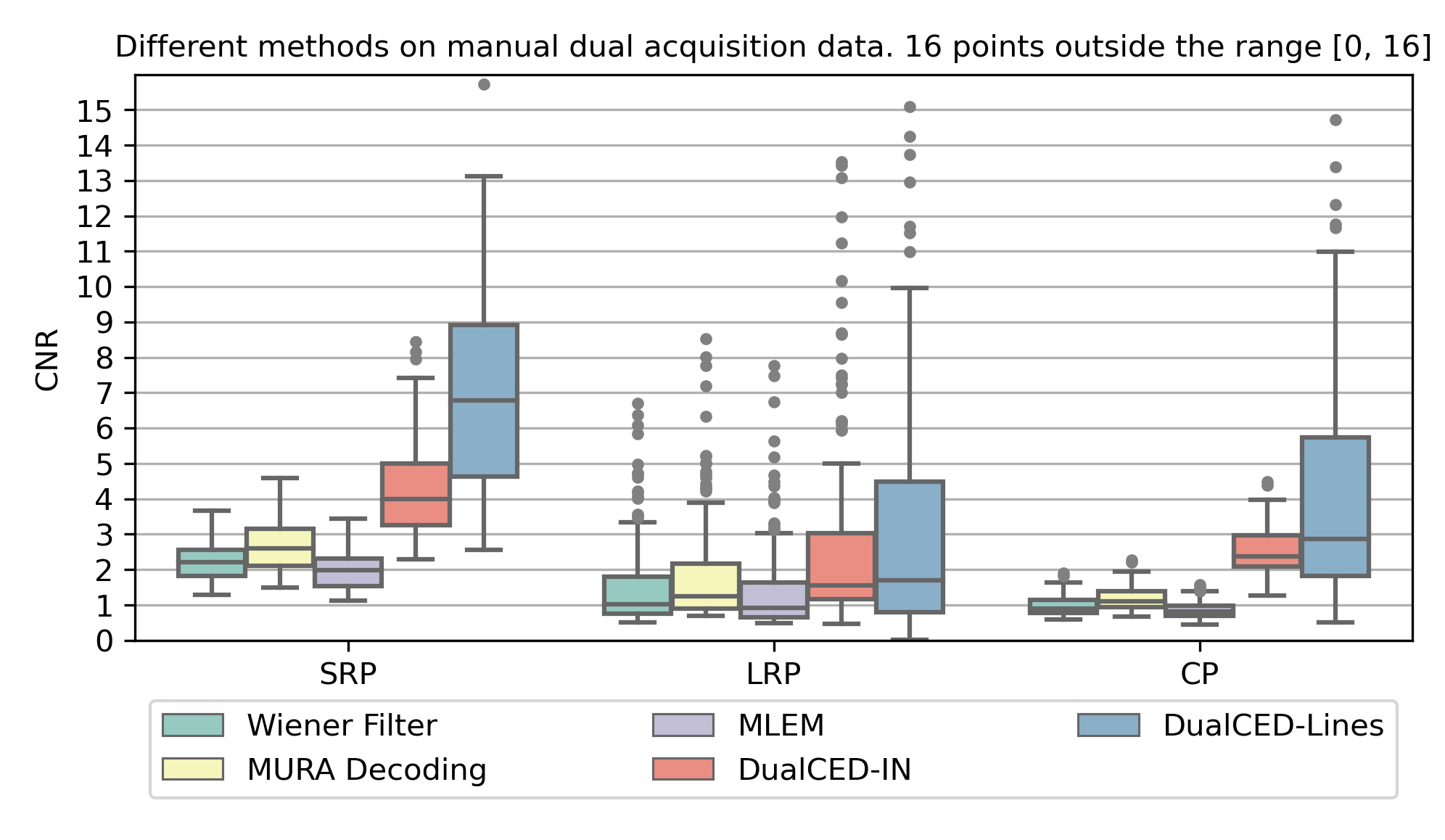}
     \caption{\label{fig:boxplot_manDual_single_acq} CNR distribution of the reconstruction of dual acquisition images separated by phantoms. 16 data points are above a CNR of 16.}
\end{figure}

\begin{figure}[bpt]
    \centering
     % trim=left bottom right top, clip
    \includegraphics[width=0.9\textwidth, trim=4 2 5 2,clip]{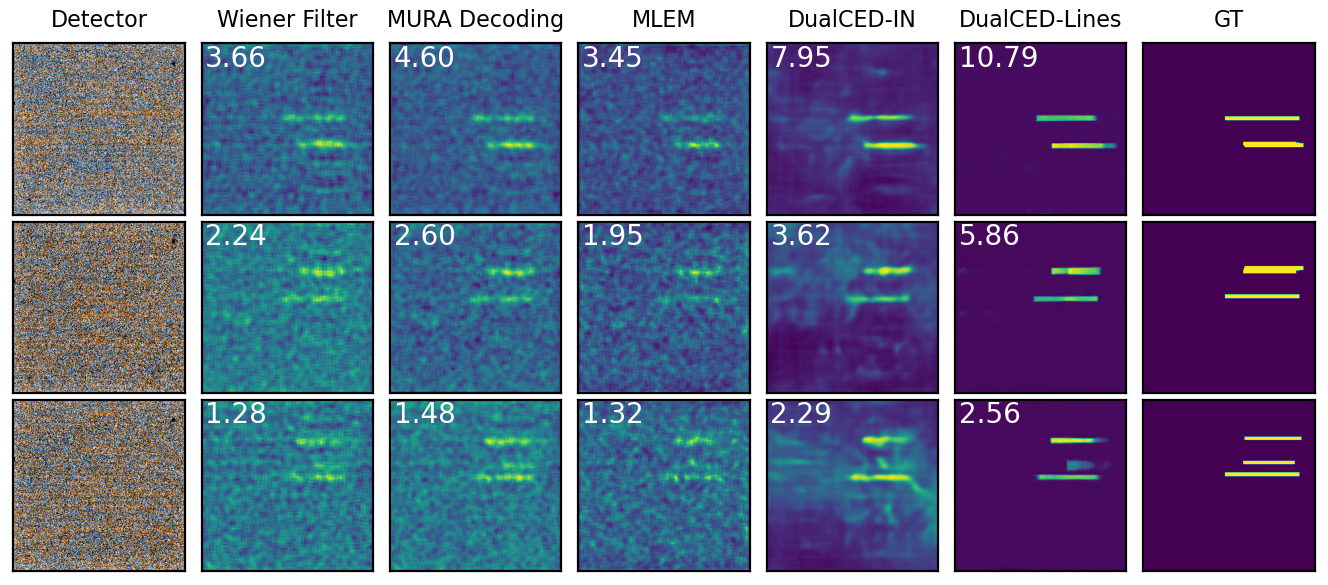}
    \includegraphics[width=0.9\textwidth, trim=4 2 5 2,clip]{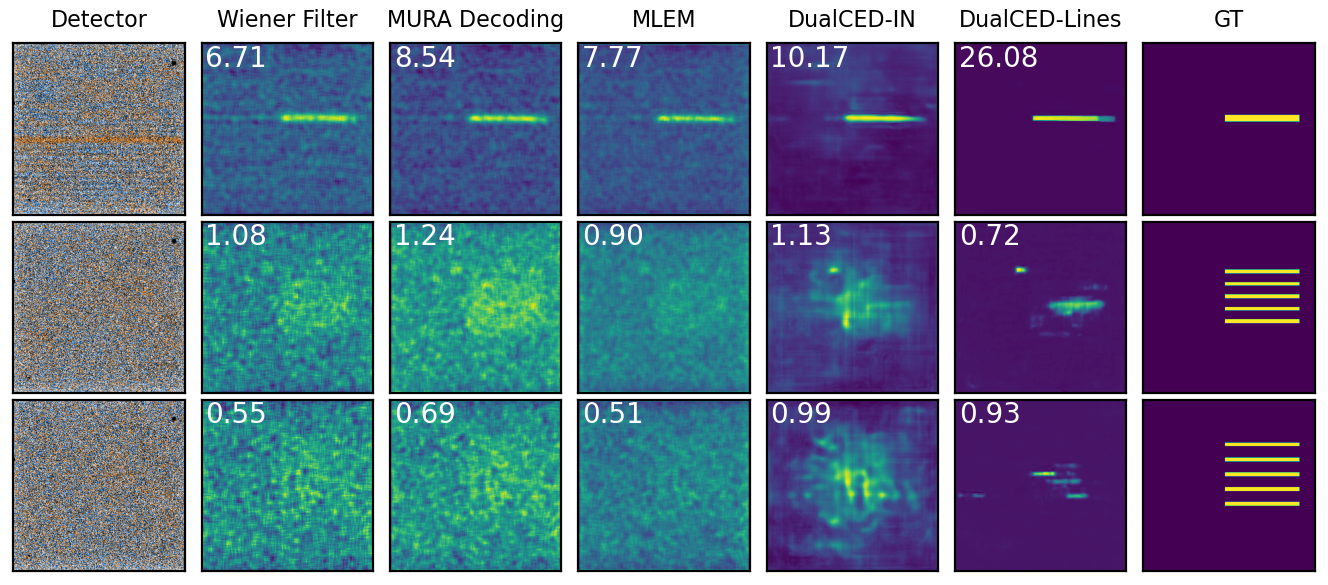}
    \includegraphics[width=0.9\textwidth, trim=4 2 5 2,clip]{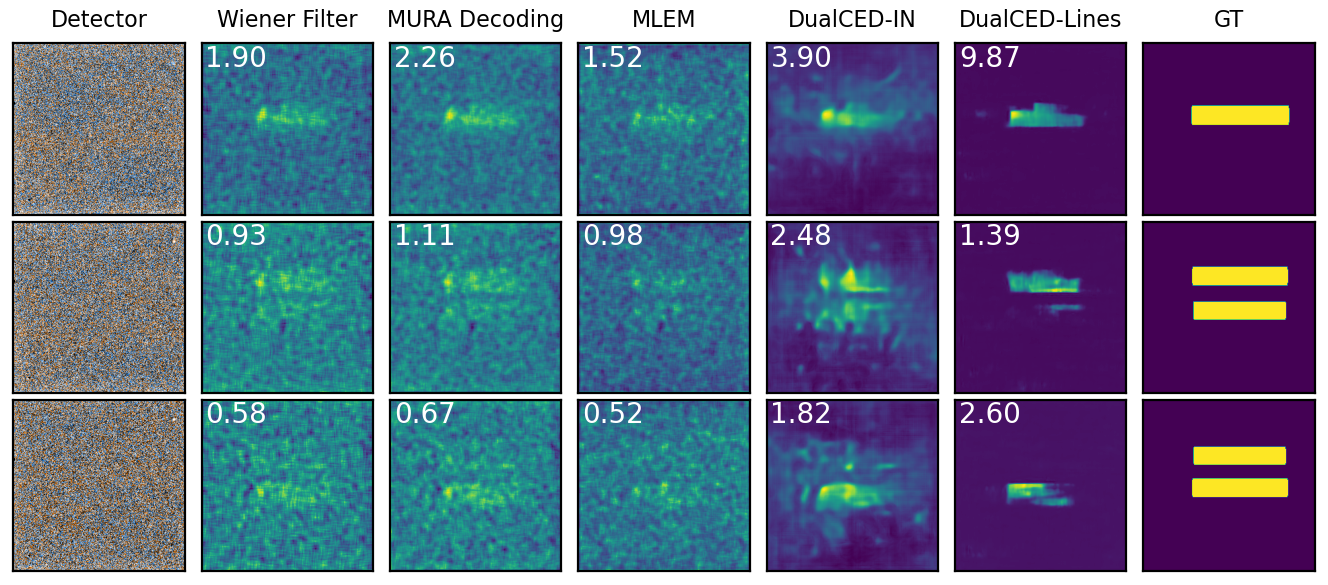}
    \caption{Exemplary images of dual image reconstructions taken from the SRP, LRP and CP with corresponding detector image and ground truth. The detector image taken with mask and with anti-mask are presented in false colors and were also histogram equalized.}
    \label{fig:BMW_overview_manDual_single_acq}
\end{figure}

The training of the DualCEDs took 18:42\,h and additionally 8:45\,h for the ImageNet based and horizontal line data set. Pairing the detector image with a second image taken with the anti-mask, resulting in dual image reconstruction, increased the reconstruction quality for almost all reconstruction methods and phantoms. Only the MLEM algorithm does not improve as much as the other methods and the CNR even decreased for the SRP by 0.12 as the boxplots in figure~\ref{fig:boxplot_manDual_single_acq} and the right half of table~\ref{tab:cnr_single_dual} show. \\
The reconstruction method with the highest median CNRs for all three phantoms are the DualCED-Lines, followed by the DualCED-IN and MURA Decoding. The CNR difference between the Wiener Filter and MURA Decoding is higher for dual image reconstruction than for the single image reconstruction. The run times presented in right half of table~\ref{tab:run_times} indicate, that only minor differences exist between the single image reconstruction methods. Mainly the MLEM algorithm run time increased by 4,910\,ms to 17,905\,ms on CPU and by 274\,ms to 1,097\,ms on the GPU.\\
As figure~\ref{fig:BMW_overview_manDual_single_acq} shows, clearer tubes and a more uniform background is noticeable in comparison to the single image reconstructions for all reconstruction methods. With the LRP in an upright position, a bright block is visible, although, single tubes remain indistinguishable. The DualCED-Lines fails to reconstruct this structure and remains dark except for a single point and a small line. Additionally, parts of the second tube of the CP become visible for all reconstruction methods, however, less prominent for the DualCED-Lines network.

\section{Discussion}
\label{sec:discussion}

\subsection{Hot-rod phantom data}
\label{sec:dis_hot_rod_phantom_data}
Even after post-processing the acquired images, noise dominates all detector images and an image structure is more distinctive for the SRP and in lesser extend for the CP and LRP. An air bubble trapped in one tube of the CP, which was found after the measurements took place, posed an additional challenge. A major source of errors in planar reconstruction is the hot-rod phantoms' depth. Even though, the maximum depth of the radiating tubes is 17\,mm and thus small in comparison to the object to camera distance of 172\,mm, one can assume that reconstruction artifacts are partially caused by this \textit{out-of-focus} effect \cite{Vassilieva2002}.\\
\noindent Despite the global outlier replacement, the acquired detector images still partially suffered from systematic pixel errors. Capturing a flood-field image would lead to an improved outlier detection and an overall higher image uniformity and is already planed for the future. Nonetheless, all reconstruction methods are equally affected by noise, the \textit{out-of-focus} effect and pixel errors, which assures a fair comparison and assessment of each method.

\subsection{Single image reconstruction}
\label{sec:single-image-reconstructions}
The most commonly used MURA Decoding proved to be good compromise between reconstruction quality and run time, despite its negligence of Poisson noise and near-field artifacts, which remain uncompensated in single image reconstruction. The Wiener Filter yields reconstructions of comparable quality but has the advantage of a more than four times faster run time. The MLEM algorithm offers higher reconstruction quality but comes with higher computational cost as it takes more than 45 times as long for a single reconstruction compared to MURA Decoding. 
It is conceivable to accelerate the run time by reducing the number of iterations. This could be achieved, for example, by combining the MLEM algorithm with a Wiener Filter for the initial guess.\\
Overall, both data-driven reconstruction methods reconstructed the phantoms better than all analytical reconstruction methods, even though the simulated training data (see Section~\ref{sec:generation_of_training_data}) did not account for near-field artifacts: Most of the reconstructions, independent from the reconstruction method, have a thin bright stripe on the top and few also at the bottom of each image. This is a hint that collimation appears caused by the small pinholes as is described in detail in \cite{Mu2006}. Especially the SRP is reconstructed 1.38 times better by CED-IN compared to the most commonly used MURA Decoding and 2.71 times better by the CED-Lines. Only in the case of particularly noisy detector images, which is the case for many of the LRP images, as explained below, the CEDs were not superior to the analytical reconstruction methods. The run time of both CEDs are almost identical and circa 286\,ms thus equally as time consuming as MURA Decoding.\\
Not only the CEDs, but all reconstruction methods have difficulties in reconstructing the phantoms when the tubes are in an upright position. In these cases, the radiating sources are spread over a large portion of the FoV. This leads to a more uniform distribution of the photons hitting the detector, which reduces the photon count per pixel. Hence, Poisson noise dominates these detector images. The almost blank reconstructions of the LRP show that the data-driven reconstruction methods cannot perform miracles and rely on a certain quality of the detector image.

\subsection{Dual image reconstruction}
\label{sec:dual-image-reconstructions}
All reconstruction methods benefit enormously from the use of the dual image approach. As has been stated before \cite{Accorsi2001, Mu2006, Vassilieva2002}, combining two captures with mask and anti-mask improves contrast and diminishes artifacts emerging from the near-field application. For example, the bright stripe at the top and bottom of the reconstructions from the single image acquisition disappeared for the dual acquisition. The MLEM algorithm did not benefit as much from the dual image approach as the other methods, suggesting that iterative forward- and backward-projection weakens the compensation effect.\\
With all methods, the background becomes more uniform and with the LRP in the upright position a bright block is even visible, although, single tubes remain indistinguishable. The DualCED-Lines fails to reconstruct this structure and only shows singular bright spots. Additionally, parts of the second tube of the CP become visible for all reconstruction methods, except for the DualCED-Lines network. Overall, however, DualCED-Lines increased its superiority over MURA Decoding with a factor of 1.45 (SRP), 1.37 (LRP) and 2.6 (CP). However, as mentioned above and depicted in figure~\ref{fig:boxplot_manDual_single_acq}, certain images are reconstructed worse than by MURA Decoding. \\
The fact that the data-driven reconstruction methods produce reconstructions with similar characteristic (e.~g. the small bright spot in the middle row of the LRP, or the highlighted tip in the top image of the CP in figure~\ref{fig:BMW_overview_manDual_single_acq}) as all analytical reconstruction methods supports the hypothesis, that the CEDs "learnt" an actual image-to-image mapping and not just searched for known patterns. \\
Unlike the data-driven CEDs, analytical reconstruction methods are based on a formal description of the imaging system. Two problems arise from this: First, it has been shown, that Convolutional Neural Networks can inherently possess instabilities \cite{Antun2020, Su2019}: For CAI, instabilities mean that perturbations of the detector image, that are caused by small unnoticeable changes of the object image, lead to nonsensical reconstructions, for example large uniform areas or negative pixel values. However, instabilities were only noticed during training and none of the final deployed networks showed instable behavior. Second, the reconstruction results are not independent from the type of training data. The different reconstruction results between DualCED-IN and DualCED-Lines can be attributed solely to the different image domains used for training. This data dependency has also been reported by \cite{Kulow2020} and indicates, that Deep Learning reconstruction methods are not capable yet, to learn a domain-independent image-to-image mapping. This can be used as an advantage as the following paragraph explains: \\
The two training sets used in this paper represent different approaches to CAI reconstruction: First, using natural photographs as object images forms a general approach aiming at the question "What do we have in front of our $\gamma$-camera?". Second, when the approximate amount, shape and size of radioactive sources is known, a training set can be tailored to that specific application. This approach contains a-priori knowledge and seeks to answers the question "Where is the source located?". For instance, when a coded aperture $\gamma$-camera is applied to the task of sentinel lymph node mapping, like several groups intend \cite{Fujii2012, Kaissas2015, Kaissas2017, Russo2020}, the radiating sources can be considered to be of spherical shape with maximally a few centimeters in diameter. A set of training data could be generated for this specific task.\\
In this paper, synthetic training data was generated (see Section~\ref{sec:generation_of_training_data}), which does not consider near-field artifacts, thus, causing a domain gap between training data and measured phantom data. Neglecting transmission or scattering of $\gamma$-photons through the mask or collimation by the small pinholes are major simplifications and inherently limit the CEDs reconstruction ability. It can be expected, that a more realistic generation of training data has the potential to improve CAI reconstruction. \\
All in all, the quantitative comparison of this paper is limited to the described experimental set-up consisting of one object-to-camera distance and a single mask with fixed thickness. To what extent the described results apply to extreme near-field application remains an open question: The more a radiation source approaches the $\gamma$-camera, the more the imaging deviates from the mathematical description as a convolution. Thus, near-field effects, mainly the angular and the collimation effect, deteriorate the detector image. Specific compensation techniques have been proposed by \cite{Accorsi2001a, Mu2006} and might be considered.

\section{Conclusion \& outlook}
\label{sec:conclusion_and_outlook}
This paper compared different reconstruction methods for planar coded aperture imaging (CAI) on data acquired by an experimental $\gamma$-camera and three hot-rod phantoms. For the given set-up, MURA Decoding, the most commonly used CAI reconstruction method, provides robust reconstructions despite the assumption of a linear system model. The Wiener Filter produces results of almost similar quality, with the advantage of being more than four times faster. For single image reconstruction, however, the MLEM algorithm reaches the highest reconstruction quality among the analytical reconstruction methods, but comes with a computing time of roughly 13\,s.\\
All in all, both Convolutional Encoder-Decoders (CED) were able to outperform analytical reconstruction methods for two out of three phantoms with single and for all three phantoms with dual image reconstruction using contrast-to-noise ratio for quality measure. The superior reconstruction quality was obtained, despite a simple low-fidelity simulation of training data. The Deep Learning approach is especially successful, when a-priori knowledge about the expected radiation sources in the form of a tailored training set is used. Even though, none of the CED reconstructions exhibited instabilities, its possibility must be kept in mind. Therefore, a more in depth analysis of the CED regarding its trustworthiness and the influence of training data is required.\\
The CED used in this paper is only one architecture among many. It would be of interest to find out how other Deep Learning architectures perform, especially with regards to the computational costs and trustworthiness.\\
For mobile applications, a short exposure time and easy handling is of importance, and thus single image acquisition is preferred. In order to keep the exposure time as short as possible, the relationship between the number of photons detected and the reconstruction quality could be investigated. By incorporating near-field artifacts and mask transmission into the generation of training data, the domain gap for single image reconstruction could be decreased and make data-driven reconstruction methods even more attractive for mobile $\gamma$-cameras.\\
\noindent The acquired hot-rod phantom data are available upon request to the co-author Vladislav Rozhkov: \href{mailto:rozhkov@jinr.ru}{rozhkov@jinr.ru}.

%\appendix
%\section{Appendix}
%\label{sec:appendix}

\acknowledgments
The authors gratefully acknowledge the data storage service SDS@hd supported by the Ministry of Science, Research and the Arts Baden-W{\"u}rttemberg (MWK) and the German Research Foundation (DFG) through grant INST 35/1314-1 FUGG and INST 35/1503-1 FUGG. This paper was partially funded by the Zentrales Innovationsprogramm Mittelstand (ZIM) under grant KK5044701BS0.

\bibliographystyle{plain} % We choose the "plain" reference style
\bibliography{references}

% \end{thebibliography}
\end{document}